\documentstyle[fleqn,prl,aps,amssymb,preprint]{revtex}

\begin{document}
\draft
\widetext

\title{Phases of two coupled Luttinger liquids}
\author{H.J. Schulz}
\address{
Laboratoire de Physique des Solides,
Universit\'{e} Paris--Sud,
91405 Orsay,
France }
\maketitle

\begin{abstract}
A model of two interacting one--dimensional fermion systems (``Luttinger
liquids'') coupled by single--particle hopping is investigated.
Bosonization allows a number of exact statements to be made. In particular,
for forward scattering only, the model contains two massless boson sectors
and an Ising type critical sector. For general interactions, there is a spin
excitation gap and either s-- or d--type pairing fluctuations dominate. It
is shown that the same behavior is also found for strong interactions. A
possible scenario for the crossover to a Fermi liquid in a many chain system
is discussed.
\end{abstract}

\narrowtext

The properties of a strictly one--dimensional interacting fermion system are
by now rather well understood.\cite{emery_revue_1d,solyom_revue_1d} The
typical phenomenology (called ``Luttinger
liquid''\cite{haldane_bosonisation}) is characterized by a separation of the
dynamics of spin and charge and by interaction--dependent power laws in many
correlation functions, and is thus quite different from Fermi liquid
behavior familiar from higher--dimensional systems. On the other hand, the
effects of coupling between parallel chains, present in any real {\em
quasi}--one--dimensional system, are still a subject of
debate.\cite{rem4,rem3,schulz_trieste} Considerable effort has been devoted
to the understanding of the properties of many coupled chains,\cite{rem3}
however, it is in many respects unclear how to connect these results to the
strictly one--dimensional case. A possible bridge between the single and
many chain cases are two (and possibly three, four, etc.) coupled
chains. The two--chain case is also of relevance for experiments on $\rm
Sr_2Cu_4O_6$, \cite{rice_srcuo} $\rm (VO)_2P_2O_7$,\cite{rem5} and possibly
the blue bronzes \cite{pouget_bronzes} (in this last case three--dimensional
phonons certainly play an important role).

The two--chain model has thus attracted considerable interest, both
analytically \cite{finkelstein_2chain,rem7,khvesh_2chain,balents_2chain} and
numerically.\cite{rem9,tsunetsugu_ladder,rem8} Nevertheless, there is little
general information on the low--lying excitations or on the possible ground
state phases. In the present paper I investigate this problem for small
interchain hopping integral and small intrachain interaction, but with their
relative size left arbitrary. Using the standard bosonization procedure, a
rather complete picture of the different possible phases and the excitation
spectrum will emerge. It will further be shown that the low--energy
properties found for weak interactions also exist in the strong--interaction
limit, suggesting that weak and strong interaction are in the same phase of
the coupled chain model.

The model I consider is given by the Hamiltonian
\begin{equation} \label{eq:h}
H=H_1 + H_2 -t_\perp \int dx (\psi^\dagger_{rs 1}
\psi^{\phantom{\dagger}}_{rs 2}+h.c.)  \;\;.
\end{equation}
Here $H_{1,2}$ are the (identical) Hamiltonians of the two
chains,\cite{emery_revue_1d,solyom_revue_1d} each characterized by a Fermi
velocity $v_F$ and forward and backward scattering interaction $g_2$ and
$g_1$, $t_\perp$ is the interchain hopping amplitude, and $\psi_{rs i}$
is the fermion field operator for right ($r=+$) or left ($r=-$) going
particles of spin $s$ on chain $i$.  To start, I neglect the backward
scattering $g_1$. The following analysis is then initially identical to that
of ref.\onlinecite{finkelstein_2chain}. The Hamiltonian is transformed by
the following steps: (i) introduce bonding and antibonding operators via
$\psi_{rs 0}=(\psi_{rs 1}+\psi_{rs 2})/\sqrt2$,
$\psi_{rs\pi}=(\psi_{rs 1}-\psi_{rs 2})/\sqrt2$; (ii)
introduce charge and spin boson fields $\phi_{\rho,\sigma;0,\pi}$
corresponding to the $0$-- and $\pi$-- fermions, following the standard
procedure; (iii) form the linear combinations $\phi_{\nu\pm} = (\phi_{\nu
0}\pm\phi_{\nu\pi})/\sqrt2$ ($\nu=\rho,\sigma$). The noninteracting
Hamiltonian (including $t_\perp$) then takes the form
\begin{equation} \label{eq:h0}
H_0 = \frac{\pi v_F}{2} \sum_{{\nu=\rho,\sigma } \atop {\alpha=\pm}}
\int dx
\left[\Pi^2_{\nu\alpha}
+\frac{1}{\pi^2}(\partial_x\phi_{\nu\alpha})^2\right] \;\;,
\end{equation}
where $\Pi_{\nu\alpha}$ is the momentum density conjugate to
$\phi_{\nu\alpha}$, and  the interaction is
\begin{eqnarray} \nonumber
\lefteqn{
H_{int,2} = \frac14 \int dx \sum_{\gamma=\pm} g^{(2)}_\gamma
\left[ \frac{1}{\pi^2} (\partial_x \phi_{\rho\gamma})^2 -
\Pi^2_{\rho\gamma} \right]}  \\
& & \label{eq:h2}
+  \frac{g^{(2)}_{00\pi\pi}}{2(\pi\alpha)^2} \int dx \cos
2 \theta_{\rho-} (\cos 2\phi_{\sigma-}+\cos 2 \theta_{\sigma-}) \;\;.
\end{eqnarray}
Here $\alpha$ is a short distance cutoff, $\partial_x\theta_{\beta\gamma}=
\pi\Pi_{\beta\gamma}$, $g^{(2)}_\gamma=
g^{(2)}_{0000} + \gamma g^{(2)}_{0\pi\pi 0}$,
and I use the notations of ref.
\onlinecite{rem7}: $g^{(2)}_{abcd}$ is the coupling constant for an
interaction scattering two particles from states $(a,b)$ into
$(d,c)$. Initially, all the $g$'s in eq. (\ref{eq:h2}) equal $g_2$, but
renormalization will give rise to differences. At energy scales higher then
$t_\perp$ an additional process of type $g^{(2)}_{0\pi0\pi}$ also exists and
is responsible for the fact that $g_2$ is not renormalized in the purely
one--dimensional problem $t_\perp =0$ (this process also only involves the
$\rho_-$ and $\sigma_-$ fields). At energies below $t_\perp$ the
$g^{(2)}_{0\pi0\pi}$-- process becomes however forbidden due to energy and
momentum conservation, and eq.(\ref{eq:h2}) is then indeed the full forward
scattering Hamiltonian.

One now can notice that the $\rho_+$ and $\sigma_+$ parts of the
Hamiltonian remain bilinear, and the corresponding fields are thus
massless. On the other hand, there are nontrivial interaction effects
for the coupled $\rho_-$ and $\sigma_-$ fields: one finds coupled
Kosterlitz--Thouless type renormalization group equations
for $g^{(2)}_{00\pi\pi}$ and $g^{(2)}_-$.
\cite{finkelstein_2chain,varma_2chain}
For the initial conditions appropriate here, these equations always scale to
strong coupling, and the standard interpretation \cite{finkelstein_2chain}
then is that there is a gap $\Delta_0 \approx t_\perp \exp(-\pi^2
v_F/|g_2|)$ for both the $\rho_-$ and $\sigma_-$ degrees of freedom.

That things are actually a bit more subtle can be seen noting that the
$\sigma_-$ part of the Hamiltonian is the continuum transfer matrix of a
two--dimensional classical XY model with twofold anisotropy field $\cos
2\phi_{\sigma-}$ (the XY spins then are ($\bbox{S}=\cos \phi_{\sigma-}$,
$\sin
\phi_{\sigma-}$)).\cite{nijs_xy,rem_majo} This model has Ising type
symmetry, with order parameter $\sin \phi_{\sigma-}$, and the symmetry of
the Hamiltonian under the duality transformation $\phi_{\sigma-}
\leftrightarrow \theta_{\sigma-}$ implies that the model is critical. The
duality symmetry is related to the fact that the left-- and right--going
fermions are independently invariant under spin rotation, i.e. there is a
chiral $SU(2) \times SU(2)$ symmetry in the fermionic model.

What are the physical properties of the pure forward scattering model?
First, there are massless modes in the $\rho_+$ and $\sigma_+$ channels,
giving a total specific heat $C(T)=(\pi T/3) (1/u_{\rho+} + 1/u_{\sigma+} +
1/2v_F)$, where the total charge and spin velocities are given by
$u_{\rho+}^2 = v_F^2-(g_2/\pi)^2$ and $u_{\sigma+} = v_F$, and the factor
$1/2$ in the last term comes from the Ising critical behavior
(with central
charge $c=1/2$ \cite{rem12}). The compressibility is determined by the
$\rho_+$ modes only and given by $\kappa^{-1} = \pi \rho_0^2 u_{\rho+}/4 K$,
where $\rho_0$ is the equilibrium particle density and $K^2 = (\pi v_F
-g_2)/(\pi v_F + g_2)$. Similarly, the (Drude) weight of the zero--frequency
peak in the conductivity is $\sigma_0 = 4 u_{\rho+} K$. As in the one--chain
case,\cite{schulz_trieste} these relations can in particular be used to
determine the coefficient $K$ which determines power laws of different
correlations functions.

Naturally, the present model does not have broken symmetry ground states,
but as in the one--chain case there are divergent susceptibilities of
different types, indicating incipient instabilities.  I first consider
$g_2>0$. To obtain the long--wavelength (low--energy) asymptotics of
correlation functions
one has to analyze the consequences of the nonlinear term in
eq.(\ref{eq:h2}) which scales to strong coupling ($g^{(2)}_{00\pi\pi}
\rightarrow \infty$). A semiclassical treatment is appropriate, and then the
energy is minimized by $\theta_{\rho-} = 0$ (there are different degenerate
solutions which all lead to identical physical results). Following standard
arguments \cite{emery_revue_1d} long--range order of the $\theta_{\rho-}$
field implies exponentially decaying $\phi_{\rho-}$ correlations. On the
other hand,
from the Ising analogy for the $\sigma_-$ sector correlations of the
order parameter
$\sin
\phi_{\sigma-}$ and its dual $\sin \theta_{\sigma-}$ then decay as
$r^{-1/4}$ whereas correlation of the non--ordering $\cos \phi_{\sigma-}$ and
$\cos \theta_{\sigma-}$ fields decay exponentially.  These points have not
been appreciated in previous work on this model. Consider now for
example
charge density oscillations which are out of phase between the two
chains, described by the operator
$O_{CDW\pi} \approx e^{i\phi_{\rho+}} \sin \phi_{\sigma+} \sin
\theta_{\sigma-}$.  From the massless modes the CDW$_\pi$ correlations then
decay as $r^{-(3+2K)/4}$, giving rise to a susceptibility diverging as
$T^{(2K-5)/4}$.  The analogous spin (SDW$_\pi$) correlations obey the
same power law, whereas in--phase correlations decay exponentially.

Similar considerations apply to BCS type instabilities. It turns out that
long--range correlations exists for the  pairing operator
\begin{equation} \label{eq:scd}
O_{SCd} = \sum_s s (\psi_{-,-s,0} \psi_{+,s,0} - \psi_{-,-s,\pi}
\psi_{+,s,\pi})
\end{equation}
and its triplet analogue. It seems appropriate to call this form ``d--wave''
because pairing amplitudes of the ``transverse'' modes $0$ and $\pi$
intervene with opposite sign. The bosonic form of this operator is given by
the same form as $O_{CDW\pi}$, with $\phi_{\rho+}
\rightarrow \theta_{\rho+}$,
$\theta_{\sigma-} \rightarrow \phi_{\sigma-}$. The corresponding
susceptibilities diverge like $T^{(2/K-5)/4}$. Because for $g_2>0$ one has
$K<1$ this divergence is subdominant compared to the CDW$_\pi$ and SDW$_\pi$
ones. It may seem surprising that the exponents do not tend to zero as $g_2
\rightarrow 0$, however one should notice that the power laws are valid for
$T<\Delta_0$, and because $\Delta\rightarrow 0$ for $g_2\rightarrow 0$ there
is a nontrivial crossover in the noninteracting limit. In all other pairing
correlations, ``s--wave'' superconductivity in particular (a plus instead of
the minus sign in eq.(\ref{eq:scd}), the leading divergent terms cancel and
one therefore has exponential decay of correlation functions and finite
susceptibilities as $T\rightarrow 0$.

For negative $g_2$ the picture changes quite drastically, because now
scaling goes to $g^{(2)}_{00\pi\pi} \rightarrow -\infty$, and consequently
the Ising order parameter is $\cos \phi_{\sigma-}$.  Now $K>1$, and the
dominant divergent susceptibility is then easily found to be standard
s--wave superconductivity, with exponent $(2/K-5)/4$. The subdominant
divergence occurs for orbital antiferromagnetic operators
\cite{nersesyan_oaf} of the form $\psi^{\dagger}_{+s\pi}
\psi^{\phantom{\dagger}}_{-s0} - \psi^{\dagger}_{+s0}
\psi^{\phantom{\dagger}}_{-s\pi}$ and its triplet analogue (the spin
nematic).

Consider now the backscattering interaction $g_1$. I will only treat the
repulsive case $g_1>0$. In a purely one--dimensional system this then scales
to zero as $g_1(E) = g_1/(1 + g_1/(\pi v_F) \ln (v_F/E \alpha))$ when the
running cutoff $E$ goes to zero. In the coupled chain problem, the
one--dimensional scaling breaks down for $E \approx t_\perp$. For small
$t_\perp$ the effective $g_1^*=g_1(t_\perp)$ will then indeed be a
perturbation. Simultaneously, $g_2$ is renormalized to
$g_2^*=g_2-g_1/2+g_1^*/2$. The backscattering Hamiltonian takes the form
\begin{eqnarray} \nonumber
H_{int,1} & = & \frac{2g_1^*}{(2\pi\alpha)^2}
\int dx \{
\cos2\phi_{\sigma+}(\cos 2\theta_{\rho-} + \cos 2\phi_{\sigma-} +
\cos 2\theta_{\sigma-}) - \cos 2\theta_{\rho-} \cos 2\theta_{\sigma-} \} \\
&&- \frac{g_1^*}{4} \int dx \left[ \frac{1}{\pi^2} \{(\partial_x
\phi_{\rho+})^2 + (\partial_x \phi_{\sigma+})^2 \} -\Pi_{\rho+}^2
-\Pi_{\sigma+}^2  \right] \label{eq:back}
\end{eqnarray}
First, the $\theta_{\rho-}$--$\theta_{\sigma-}$ interaction now breaks the
self--duality of the $\phi_{\sigma-}$ fields. As $\cos 2\theta_{\rho-}$
already has a nonzero expectation value from the $g_2$ interaction, one now
also finds a gap in the $\sigma_-$ modes, of order
$\Delta_\sigma=(g_1^*/g_2^*) \Delta_0$. In the Ising model language, this
corresponds to a deviation from criticality, long--range order and
exponentially decaying $\sin \theta_{\sigma-}$ correlations.  Secondly, the
forward scattering interaction also leads to a nonzero expectation value of
$\cos 2\theta_{\rho-} + \cos 2\phi_{\sigma-} + \cos 2\theta_{\sigma-}$ which
by spin rotation invariance has to be positive. The leading order effect of
the first term in eq.(\ref{eq:back}) then is to open a gap also in the
$\sigma_+$ degrees of freedom, given, up to numerical factors, by
$\Delta_\sigma$. In the presence of the backscattering interaction {\em
there thus is a gap in all the magnetic excitations}.

In correlation functions, to leading order one now replaces $\phi_{\sigma+}$
by its classical value $\pi/2$. One then finds for $g_2>0$ a decay of the
SC$d$ correlations as $r^{-1/2K}$, giving rise to a divergence of the
corresponding susceptibility as $T^{1/2K-2}$, where now $K^2=(\pi
v_F-g_2+g_1/2)/(\pi v_F+g_2-g_1/2)$. On the other hand the CDW$_\pi$ and
SDW$_\pi$ operators contains the Ising disorder field, and therefore these
correlations decay exponentially. A divergent density response exists for
correlations of the form $\langle [O_{CDW\pi}(r)]^2 [O_{CDW\pi}(0)]^2
\rangle \propto \cos (2(k_{F0}+k_{F\pi})r) r^{-2K}$,\cite{balents_2chain}
because here the operator $\sin^2\theta_{\sigma-} \approx 1/2$
appears. Perturbative and symmetry arguments show that the same
contribution also exists in the density correlations: $\langle
n(r) n(0)\rangle \propto
\cos (2(k_{F0}+k_{F\pi})r) r^{-2K}$, in analogy with the $4k_F$
oscillations of a single chain. However at least for weak interactions
($K\rightarrow
1$)
the corresponding susceptibility is much weaker than the $SCd$ pairing,
i.e. {\em the two--chain model has predominant pairing fluctuations even for
purely repulsive interactions}.\cite{rem6} In the regime of negative $g_2$
the leading divergent susceptibility is s--wave superconductivity with
exponent $1/2K -2$. The precise boundary between the two regimes can be
determined from the scaling equations of ref.\onlinecite{varma_2chain} and
is given by $g_1=2g_2$. The triplet susceptibilities (spin density wave or
triplet superconductivity) are suppressed by the spin gap. The spin gap
gives rise to ``anomalous flux quantization'',\cite{rem13} and there is
also a gap for single--particle excitations.

The power laws discussed above apply in the temperature region below
$\Delta_0$. In the intermediate region $\Delta_0 < T < t_\perp$ the
$g^{(2)}_{00\pi\pi}$ term in eq.(\ref{eq:h2}) has little effect, and one
then can obtain the temperature dependence of different correlation
functions from a purely bilinear Hamiltonian. For example, for CDW$_0$
susceptibilities one finds a power law $T^{(K-1)/2}$, whereas in the
one--dimensional region $T > t_\perp$ one has a behavior as $T^{K-1}$. The
important point here is that in the intermediate region the interaction
dependent exponent is smaller than in the high--temperature region, i.e.
below $t_\perp$ the system behaves more closely like a Fermi liquid than at
high temperatures.

Let me now briefly consider the strongly interacting case. For sufficiently
strong intrachain interactions, i.e. small parameter $K_\rho$ of the
individual chains, single--particle hopping is renormalized to zero, however
simultaneously particle--hole tunneling processes
appear.\cite{rem3,schulz_trieste} Introducing $\phi_{\nu\pm} =(\phi_{\nu 1}
\pm \phi_{\nu 2})/\sqrt2$, where $\phi_{\nu 1,2}$ are the boson fields of
the individual chains, for the purely forward scattering case, these terms
take the form $J \cos 2 \phi_{\rho-} (\cos 2 \phi_{\sigma-} + \cos 2
\theta_{\sigma-})$. One again has a duality symmetry, $\phi_{\sigma-}
\leftrightarrow \theta_{\sigma-}$, and the same types of power-law
correlations as in the weak--coupling case appear. Introducing now
intrachain backscattering, the duality symmetry is broken and, again as in
the weak--coupling case, only SCd correlations (exponent $1/2K_\rho$) and
$4k_F$ charge correlations (exponent $2K_\rho$) remain. The types of
possibly divergent response functions and the scaling relations between
different exponents are thus identical for weak and strong interaction.
This strongly suggests that {\em this type of behavior actually holds for
arbitrary interaction strength}. Note that the density correlations decay
more slowly than the pairing correlations only for $K_\rho<1/2$. This
typically corresponds to rather strong repulsion: for example, in the
one--dimensional Hubbard model one reaches $K_\rho=1/2$ only for infinite
repulsion.\cite{schulz_trieste} Another interesting strong--coupling model
is the ``t--J ladder''.\cite{tsunetsugu_ladder} Here in the limit of strong
interchain exchange a mapping onto an effective single--chain hard core
boson model can be made, leading again to the same low--energy properties as
in the weak--coupling limit.\cite{rem10} Recent numerical results
\cite{rem9,rem8} confirm this point.

The exponents $K-1$ and $(K-1)/2$ valid for the single and double chain
problems suggest that for $N$ chains coupled by near--neighbor interchain
hopping one might have an anomalous exponent $(K-1)/N$ at $T<t_\perp$. To
see how such a behavior can possibly arise, in analogy to the two--chain
case one can go to momentum space in the transverse direction. The
noninteracting bosonized Hamiltonian then is
\begin{equation} \label{eq:hn}
H_0 = \frac{\pi v_F}{2} \sum_{{\nu=\rho,\sigma} \atop {k_\perp}}
\int dx
\left[\Pi^2_{\nu k_\perp}
+\frac{1}{\pi^2}(\partial_x\phi_{\nu k_\perp})^2\right] \;\;,
\end{equation}
Following standard arguments \cite{shankar_revue} I now only consider
forward scattering interactions which for states at the Fermi energy are
consistent with both energy and momentum conservation. The analogue of the
first term in eq.(\ref{eq:h2}) then is
\begin{equation} \label{eq:h2m}
H_{\text{int,2}} = \frac{g_2}{2} \int dx
\left[ \frac{1}{\pi^2} (\partial_x \phi_{\rho 0})^2 -
\Pi^2_{\rho 0} \right] \;\;,
\end{equation}
where $\phi_{\rho x}$ is the Fourier transform of $\phi_{\rho k_\perp}$
with respect to $k_\perp$. The
important point here is that {\em only the mode at $x=0$ is affected by the
interactions}. A standard calculation then leads to a decay of $CDW$
correlations as $r^{-2-(K-1)/N}$, giving rise to a susceptibility behaving
as $T^{(K-1)/N}$. Similarly, the single particle Green function decays as
$r^{-1-\delta}$, with $\delta=(K+1/K-2)/4N$, leading to a singularity of the
momentum distribution function as $|k-k_F|^{\delta}$.\cite{rem11} In the
limit of a large number of coupled chains the anomalous exponents now
vanish, and in particular one recovers a Fermi liquid like momentum
distribution function in this description.

Clearly, a number of interactions has been neglected in this argument. First
there are Cooper type ($(k,-k)\rightarrow (k',-k')$) and possibly nesting
interactions, the prototype of which is given by the $g^{(2)}_{00\pi\pi}$
term in eq.(\ref{eq:h2}). By analogy with that case I expect these
interactions to give rise to a gap of order $\Delta_0$, and to ordered
ground states for $N\rightarrow \infty$. Thus the power laws of the
preceding paragraphs are valid in the temperature region
$\Delta_0<T<t_\perp$.  Moreover, there are interactions that involve at
least one state not exactly at the Fermi energy. Though these interactions
cannot directly affect the low--energy physics, they in general will lead to
renormalizations of $g_2$. The above arguments remain valid if these
renormalizations are nonsingular. To which extent this is correct is
currently under investigation.

In conclusion, I've investigated the phase diagram and excitation spectrum
of two Luttinger liquids coupled by single--particle hopping, and proposed a
possible extension to many coupled chains. The conclusions are valid for
small hopping amplitude, but the same types of divergent responses (d--type
superconductivity and $4k_F$ charge density in the case of repulsion) occur
for both weak and strong interactions, suggesting that this type of behavior
is to be found for rather general interactions.  The fact that for strong
interactions interaction interchain hopping renormalizes to zero
\cite{schulz_trieste,rem3,rem7} only affects properties at intermediate
energy scales (above the spin gap). Contrary to the case of a single
chain, the pure forward--scattering model is found to be a singular
line in the phase diagram, with Ising type criticality.

I am grateful to L. Balents, T. Einarsson, M.P.A. Fisher, T. Giamarchi,
R. Noack, D. Poilblanc, J.P. Pouget, and H. Tsunetsugu for stimulating
discussions. Laboratoire de Physique des Solides is a Laboratoire Associ\'e
au CNRS.


\begin{thebibliography}{10}
\bibitem{rem_majo}
The same conclusion is reached going back to a fermionic representation in the
  ($\rho_-$,$\sigma_-$) sector. Then a massless Majorana fermion mode
  decouples, a well--known representation of the critical two--dimensional
  Ising model.
\bibitem{rem6}
In a more realistic model, the Fermi velocities $v_{F0,\pi}$ of the two bands
  are different. This can be shown to lead to effective parameters $u_{\rho+}$
  and $K$ given by $(u_{\rho+} K)_{eff} = u_{\rho+} K$, $(u_{\rho+}/ K)_{eff} =
  u_{\rho+}/ K - (v_{F0}-v_{F\pi})^2/4\overline{v_F}$.
\end{thebibliography}
\end{document}